\documentclass[sigconf]{acmart}
\usepackage{comment} % for multiline comment
\usepackage{multirow} % for table
\usepackage{makecell}

\usepackage{amssymb}
\usepackage{colortbl}
\usepackage{framed}

\copyrightyear{2026}
\acmYear{2026}
\setcopyright{acmlicensed}
\setcctype{by}
\acmConference[EASE 2026]{The 30th International Conference on Evaluation and Assessment in Software Engineering}{9–12 June, 2026}{Glasgow, Scotland, United Kingdom}
\acmPrice{}

\begin{document}
%\title{Title1}
\title{An Empirical Study of Database Security Topics on Technical Social Forums of Software Developers}
\title{REStack: A Large-Scale Dataset of Reverse Engineering Discussions from Stack Exchange}
% Challanges among database security developers: large scale empirical study on Stack Overflow
% Technical challanges faced by database security developers

%\titlenote{Produces the permission block, and copyright information}
%\subtitlenote{The full version of the author's guide is available as

\renewcommand{\shorttitle}{A Large-Scale Dataset of Reverse Engineering Discussions from Stack Exchange}

\author{ Md Humaun Kabir, Md Rakibul Islam, Farha Kamal} 
\affiliation{
  \institution{Lamar University}
  \city{Beaumont}
  \state{Texas}
  \country{USA}
}
\email{{mkabir13, mislam108, fkamal}@lamar.edu}

\renewcommand{\shortauthors}{Islam et al.}

% \author{Anonymous Author(s)*}

% % The default list of authors is too long for headers}
% \renewcommand{\shortauthors}{Anonymous Author(s).}

\begin{abstract} 

Reverse engineering (RE) is a critical activity in software engineering and cybersecurity, supporting tasks such as malware analysis, vulnerability discovery, legacy system maintenance, and firmware inspection. Despite its importance, there is limited empirical understanding of the challenges, topics, and knowledge gaps faced by RE practitioners in real-world settings, and no publicly available dataset has systematically captured RE discussions from developer Q\&A forums. 

In this paper, we present \texttt{REStack}, a large-scale dataset of RE discussions collected from Stack Overflow and the dedicated Reverse Engineering Stack Exchange site. The dataset comprises over 12,000 RE-related posts spanning more than 15 years. Using Latent Dirichlet Allocation (LDA) with Genetic Algorithm (GA)–based hyperparameter optimization, followed by manual topic labeling, we identify 23 semantically coherent RE topics grouped into six high-level thematic categories. The dataset is further enriched with metadata and difficulty indicators derived from community interaction signals, such as unanswered rates and response times.

Our analysis reveals that RE discussions are predominantly practical and task-oriented, with strong emphasis on debugging, decompilation, and system-level analysis, while topics related to memory, firmware, and file format analysis exhibit high difficulty and unresolved rates. Beyond empirical characterization, \texttt{REStack} provides a reusable resource for empirical studies, educational research, and the development and evaluation of AI- and LLM-based developer assistance tools for RE. By releasing the dataset and accompanying scripts, this work aims to facilitate reproducible research and advance data-driven support for RE practice.
\end{abstract}

\begin{CCSXML}
<ccs2012>
   <concept>
       <concept_id>10002978.10003022.10003465</concept_id>
       <concept_desc>Security and privacy~Software reverse engineering</concept_desc>
       <concept_significance>500</concept_significance>
       </concept>
 </ccs2012>
\end{CCSXML}

\ccsdesc[500]{Security and privacy~Software reverse engineering}


\keywords{Reverse engineering, Stack Exchange, Topics,
Dataset}

\maketitle

% \begin{IEEEkeywords}
% topic models, empirical study, secure software development.
% \end{IEEEkeywords}

% Index
% 1 Introduction
% 2 Methodology
% 3 Case Study Results
% 3.1 RQ1: What topics Database Security
% developers asking about?
% 3.2 RQ2: What types of questions are Database
% Security developers asking about?
% 3.3 RQ3: Which topics are the most difficult to answer?
% 4 Discussion & Implications
% 4.1 Topics Evolution
% 4.2 compare to other SE Fields
% 4.3 Implications
% 5 Related Work
% 6 Threats to Validity
% 7 Conclusion
%
%
% 

\section{Introduction}

Reverse engineering (RE) is a fundamental practice in software engineering and cybersecurity, enabling developers and researchers to analyze software and hardware systems in the absence of source code. RE plays a critical role in a wide range of activities, including malware analysis, vulnerability discovery, legacy system maintenance, software interoperability, and firmware inspection. As software systems continue to grow in complexity and security mechanisms become increasingly sophisticated, the demand for effective RE techniques and tools has intensified.

Despite its importance, RE remains a technically challenging and expertise-intensive activity. Practitioners frequently rely on community-driven question-and-answer (Q\&A) platforms, such as Stack Overflow (SO)~\cite{StackOverFlow} and the Reverse Engineering Stack Exchange site~\cite{RESite}, to seek guidance on tools, techniques, and problem-solving strategies. These platforms contain a rich record of real-world RE challenges, capturing not only what developers struggle with, but also how knowledge is shared, reused, and sometimes left unresolved. However, unlike other software engineering domains, RE has received relatively limited attention in large-scale empirical studies, and no publicly available dataset has systematically curated and analyzed RE discussions across developer forums.

Prior research has demonstrated the value of mining SO data to understand developer challenges in areas such as security, concurrency, mobile development, big data, and refactoring~\cite{yang2016security,ahmed2018concurrency,rosen2016mobile,GoingBig}. While these studies provide important insights, they either treat RE as a minor subtopic or exclude it altogether. Consequently, there is a lack of domain-specific datasets that capture the unique characteristics of RE, including low-level program analysis, binary reasoning, firmware inspection, and anti-debugging techniques. This gap limits the ability of researchers to conduct reproducible empirical studies, develop RE-focused educational resources, or evaluate AI- and LLM-based developer support tools in this domain.

To address this gap, we present \texttt{REStack}, a large-scale dataset of RE discussions collected from SO~\cite{StackOverFlow} and the dedicated RE Stack Exchange site~\cite{RESite}. The dataset comprises over 12,000 RE-related posts spanning more than 15 years of community activity. Using a combination of Latent Dirichlet Allocation (LDA)~\cite{LDA} with Genetic Algorithm (GA)~\cite{Alibrahim2021}–based hyperparameter optimization and manual topic labeling, we identify 23 semantically coherent RE topics grouped into six high-level thematic categories. In addition to topic annotations, the dataset includes rich metadata and difficulty indicators derived from community interaction signals, enabling multifaceted analysis of popularity, complexity, and knowledge gaps.

The contributions of this paper are threefold. First, we introduce the first publicly available, curated dataset dedicated exclusively to RE discussions from Stack Exchange. Second, we provide a structured topic taxonomy and quantitative characterization of RE topics, offering insights into the breadth and depth of challenges faced by practitioners. Third, we demonstrate the utility of the dataset by analyzing topic distributions and difficulty patterns, and by outlining research opportunities in areas such as automated question answering, expert recommendation, RE education, and benchmarking large language models for low-level software reasoning.

By releasing \texttt{REStack} and the associated scripts~\cite{Replication_Package}, this work aims to support reproducible research and foster further studies at the intersection of RE, software engineering, and artificial intelligence. We envision the dataset serving as a foundational resource for researchers, educators, and tool developers seeking to better understand and support RE practice.

\section{Dataset Construction Methodology}\label{secMethod}

\noindent\textbf{Data Sources.}
To gather developer discussions on RE, we collected data from two technical Q\&A platforms within the Stack Exchange network~\cite{StackExchangeURL}: the dedicated RE site~\cite{RESite} and SO~\cite{StackOverFlow}, the largest Stack Exchange platform for general software development discourse. Both platforms structure their content as question-and-answer posts, each carrying metadata such as a unique identifier, post type (question or answer), title, body, tags, creation date, view count, and score. Questions additionally store the identifier of their accepted answer, which is designated by the original poster. Tags are short descriptors summarizing a question's subject matter, and a single question may carry multiple tags.

\noindent\textbf{Identifying RE-Related Tags in SO.}
Since all posts on the RE site are inherently relevant to the domain, no filtering is required there. SO, however, spans a broad range of software engineering topics, so we must isolate posts pertaining to RE. We do this by first identifying tags associated with RE content.

To build a comprehensive tag set, we follow an approach inspired by prior work~\cite{ahmad2020chatbot, uddin2021iot}. Starting with all posts explicitly tagged \texttt{reverse-engineering}, we obtain an initial seed set of 1,941 posts, deliberately avoiding any additional tags at this stage to prevent noise from contaminating the foundation for co-occurring tag discovery. We then extract every tag that appears alongside the seed tag \texttt{reverse-engineering} in these initial posts and evaluate each candidate tag using two heuristics established in the literature~\cite{ahmad2020chatbot, uddin2021iot}: the \textit{Tag Relevance Threshold} (TRT) and the \textit{Tag Significance Threshold} (TST).

The \textbf{TRT} measures the fraction of posts carrying a given tag that are also tagged with \texttt{reverse-engineering}:
\begin{equation}
\text{TRT} = \frac{\text{No.\ of RE posts for the tag}}{\text{Total no.\ of posts for the tag in SO}}
\end{equation}

A tag with a TRT of 0.21, for instance, indicates that 21\% of its posts are RE-related. TRT alone, however, can be misleading for tags with very few posts: a tag appearing in only three posts would achieve a TRT of 33.3\% from a single RE-related post, potentially inflating the relevance of peripheral tags.

To counteract this, we apply the \textbf{TST}, which captures a tag's prominence within the RE corpus:
\begin{equation}
\text{TST} = \frac{\text{No.\ of RE posts for the tag}}{\text{No.\ of posts for the most frequent RE tag}}
\end{equation}

A TST of 0.03 for a tag $\mathcal{T}$, for example, indicates it accounts for 3\% of all posts tagged with \texttt{reverse-engineering}, helping to distinguish genuinely recurrent RE topics from incidental ones.

Only tags exceeding the minimum thresholds for \emph{both} metrics were retained. The first two authors, each with more than three years of experience in RE, independently reviewed candidate tags across multiple threshold combinations, sampling and examining representative posts to pinpoint where tags began to deviate from core RE subject matter. Following thorough review and consensus discussions, we set the TRT threshold at 11\% and the TST threshold at 0.12\%, values that optimally balance coverage of RE-related content against inclusion of unrelated noise and are consistent with thresholds adopted in comparable studies~\cite{ahmad2020chatbot, uddin2021iot}. Exactly four additional tags satisfied both criteria: \texttt{disassembly}, \texttt{decompiling}, \texttt{ollydbg}, and \texttt{ida}. Combined with \texttt{reverse-engineering}, this yields five tags used to retrieve RE-related posts from SO.

\noindent\textbf{Extracting RE-Related Posts from RE and SO.}
Posts from the RE site were retrieved through its query interface\footnote{\url{https://data.stackexchange.com/reverseengineering/queries}} via an SQL query, yielding all 9,845 available posts. For SO, we queried the platform's data dump stored in Google BigQuery\footnote{\url{https://cloud.google.com/bigquery}} and selected all posts carrying at least one of the five RE-related tags. This produced 2,448 questions. Combining both sources, we obtained a total of 12,293 posts, retrieved on May 7, 2025.

\noindent\textbf{Preprocessing RE Posts.}
We apply a systematic preprocessing pipeline to improve the clarity and semantic quality of the corpus. Code blocks (enclosed in \texttt{<code>} tags), HTML markup (e.g., \texttt{<p>}, \texttt{</p>}), and hyperlinks are removed. Numeric characters and punctuation are discarded, as they contribute little to semantic interpretation. Common English stopwords (e.g., ``a'', ``can'') are eliminated using the NLTK stopwords corpus~\cite{NLTKSTopWords} to reduce noise in subsequent topic modeling. Finally, lemmatization is applied to reduce inflected or derived words to their canonical forms (e.g., ``going'' becomes ``go''), yielding a normalized corpus suitable for topic modeling. For each post, we construct an additional textual feature, \emph{topicData}, by concatenating the post's title and body into a single text sequence, which serves as input to the topic modeling framework described below.

\noindent\textbf{Determining the Optimal Number of Topics.}
To uncover the themes most commonly discussed by RE practitioners, we apply Latent Dirichlet Allocation (LDA)~\cite{LDA}, a widely used unsupervised algorithm for thematic clustering of textual data~\cite{openja2020release, CI, yang2016security, Refactor, ahmad2020chatbot}. LDA performance is highly sensitive to hyperparameter settings~\cite{CI, Refactor, yang2016security}: too few topics produce overly broad themes, while too many introduce redundancy and semantic overlap. Beyond the number of topics, parameters such as iteration count, chunk size (documents processed per batch), and passes (complete corpus traversals) critically affect model convergence and coherence, and their interdependencies make manual tuning infeasible~\cite{Li2021, Yang2020}.

We therefore employ a Genetic Algorithm (GA)~\cite{holland1992adaptation} to automate hyperparameter optimization, a strategy well-suited to multi-dimensional search problems in software engineering~\cite{CI, Saidani2022, Alibrahim2021}. The GA tunes four parameters: number of topics (2--50), iterations (10--5,000), chunk size (10--2,000), and passes (1--100), using Topic Coherence~\cite{CI, TopicCoherence} as the fitness metric. All experiments are implemented with the \texttt{Gensim} library~\cite{Gensim}. The resulting optimized model comprises 23 semantically coherent RE topics, trained over 2,000 iterations with a chunk size of 500 and 200 passes.

\noindent\textbf{Manual Topic Labeling.}
LDA produces topics as anonymous, numerically indexed clusters, so assigning interpretable labels is essential. We use an open card-sorting approach~\cite{CardSorting}, a qualitative labeling method widely adopted in the literature~\cite{ahmad2020chatbot, ahmed2018concurrency, GoingBig, rosen2016mobile, yang2016security} that allows thematic labels to emerge inductively from the data rather than from a predefined taxonomy. Each of the first two authors independently reviewed the top 20 keywords for every topic and read a random sample of at least 30 representative posts per topic (or all posts for topics with fewer than 30). Each annotator then proposed a concise label capturing the topic's central theme. 

To assess the reliability of the manual topic labeling process, we computed 
Cohen's kappa ($\kappa$)~\cite{cohen1988} as a measure of inter-rater agreement 
between the two annotators. Cohen's kappa is preferred over raw percentage 
agreement because it corrects for the probability of agreement occurring by 
chance, making it a more conservative and defensible measure of labeling 
consistency~\cite{landis1977}. The 
two annotators achieved an inter-rater agreement of $\kappa=$\textbf{0.73}, 
which falls within the \textit{substantial} agreement range 
according to the widely adopted scale proposed by Landis and 
Koch~\cite{landis1977}. In cases 
of disagreement, the annotators engaged in structured discussion until consensus 
was reached, after which all 23 topic labels were finalized. This inter-rater agreement protocol is consistent 
with that adopted in comparable SO mining 
studies~\cite{yang2016security, GoingBig,ahmad2020chatbot}. In addition, to enable higher-level analysis, the labeled topics were subsequently grouped into broader thematic categories based on content similarity by the annotators together.

\begin{table}[h]
\caption{REStack dataset summary: structural characteristics 
and community engagement statistics.}
\label{tab:dataset_summary}
\small
\begin{tabular}{llr}
\toprule
\textbf{Dimension} & \textbf{Metric} & \textbf{Value} \\
\midrule

\multirow{7}{*}{\textbf{Structure}}
  & Total posts                        & 12,293        \\
  & \quad From RESE                    & 9,845 (80.1\%)  \\
  & \quad From SO          & 2,449 (19.9\%)  \\
  & Temporal coverage                  & Aug 2008 -- Apr 2025 \\
  & LDA topics                         & 23            \\
  & Thematic categories                & 6             \\
  % & Unique SO tags                     & 429           \\
\midrule

\multirow{4}{*}{\textbf{Resolution}}
  & Posts with accepted answer         & 5,470 (44.5\%) \\
  & Posts without accepted answer      & 6,824 (55.5\%) \\
  & Posts receiving zero answers       & 2,513 (20.4\%) \\
  & Posts receiving $\geq$1 answer     & 9,781 (79.6\%) \\
\midrule

\multirow{4}{*}{\textbf{Score}}
  & Mean score                         & 2.68          \\
  & Median score                       & 1.0           \\
  & Minimum score                      & $-$9          \\
  & Maximum score                      & 501           \\
\midrule

\multirow{4}{*}{\textbf{View Count}}
  & Mean view count                    & 2,378.9       \\
  & Median view count                  & 562           \\
  & Minimum view count                 & 7             \\
  & Maximum view count                 & 424,956       \\
\midrule

\multirow{3}{*}{\textbf{Comments}}
  & Mean comments per post             & 1.74          \\
  & Minimum comments                   & 0             \\
  & Maximum comments                   & 54            \\
\midrule

\multirow{3}{*}{\textbf{Favourites}}
  & Posts with at least one favourite  & 672 (5.5\%)   \\
  & Mean favourites (all posts)        & 4.69          \\
  & Maximum favourites                 & 314           \\

\bottomrule
\multicolumn{3}{l}{\footnotesize RESE = Reverse Engineering 
Stack Exchange.}\\
\multicolumn{3}{l}{\footnotesize Score = community vote 
total (upvotes $-$ downvotes).}\\
\end{tabular}
\end{table}

\begin{table}[htbp]
\centering
\caption{Summary of Reverse Engineering Topics}
\vspace{-0.3cm}
\label{TblTopics}
\resizebox{\columnwidth}{!}{%
\begin{tabular}{@{}lll@{}}
\toprule
\textbf{Topic Category} & \textbf{Topic} & \textbf{Number} \\
\midrule
\multirow{3}{*}{\begin{tabular}[c]{@{}l@{}}Database \& \\Software Architecture RE\end{tabular}} 
& Data Encoding \& ORM Issues & 1,156 \\
& Database Schema Reverse Engineering (ORM Tools) & 43 \\
& Software Analysis Methods (Modeling, GUI, Memory) & 114 \\
\midrule
\multirow{6}{*}{\begin{tabular}[c]{@{}l@{}}Debugging \& \\Anti-Debugging Techniques\end{tabular}}
& 64-bit Compatibility \& Bitwise Operations & 125 \\
& Advanced Debugging Techniques & 242 \\
& Advanced Tools \& Legacy Debugging & 96 \\
& Anti-Debugging \& Obfuscation Challenges & 16 \\
& Assembly Code Analysis \& Automation & 159 \\
& Executable Patching \& Debugger Issues & 470 \\
\midrule
\multirow{4}{*}{\begin{tabular}[c]{@{}l@{}}Decompilation, Disassembly, \\\& Code Understanding\end{tabular}}
& Decompilation \& Algorithm Reconstruction & 481 \\
& Function Entry/Exit Manipulation & 399 \\
& IDA Pro \& High-Level Code Reversing & 80 \\
& OS Kernel \& Software Reverse Engineering & 147 \\
\midrule
\multirow{2}{*}{Games, Challenges, \& Educational RE}
& Game Hacking \& Miscellaneous Analysis & 114 \\
& Reverse Engineering Challenges (Games \& Puzzles) & 3,932 \\
\midrule
\multirow{3}{*}{\begin{tabular}[c]{@{}l@{}}Memory, Firmware \\\& System-Level Analysis \end{tabular}}
& File Formats \& System Behavior & 807 \\
& Hardware Hacking \& Device Emulation & 2,056 \\
& Memory, Firmware \& File Format Analysis & 68 \\
\midrule
\multirow{5}{*}{RE Fundamentals}
& Basic Assembly Reversing \& Debugging & 223 \\
& Data Decoding \& System Security & 240 \\
& Function Call Tracing \& Hooking & 1,058 \\
& General Data Analysis \& Reverse Engineering Tools & 220 \\
& General Reverse Engineering Challenges & 47 \\
\bottomrule
\end{tabular}%
}
\vspace{-0.6cm}
\end{table}

\section{Dataset Description}

Table~\ref{tab:dataset_summary} consolidates the structural 
characteristics and community engagement statistics of REStack. 
The dataset comprises 12,293 posts drawn from two complementary 
sources: 9,845 posts (80.1\%) from the dedicated Reverse 
Engineering Stack Exchange site (RESE) and 2,449 posts (19.9\%) 
from SO, spanning approximately 17 years of 
community activity from August 2008 to April 2025.

In terms of resolution, 5,470 posts (44.5\%) carry an accepted 
answer designated by the original question author, while the 
remaining 6,824 (55.5\%) remain unresolved. More strikingly, 
2,513 posts (20.4\%) received no community response whatsoever, 
meaning that one in five RE questions attracted zero answers — 
a figure that underscores the depth of knowledge gaps in the 
RE community and motivates the need for AI-assisted tooling 
and improved documentation.

Community engagement signals exhibit a high degree of 
right-skew across all dimensions. Post scores range from 
$-9$ to $501$ with a mean of $2.68$ and a median of only 
$1.0$, indicating that the majority of posts receive modest 
engagement while a small number of highly upvoted questions 
attract disproportionate community attention. View counts 
similarly span several orders of magnitude, from a minimum 
of 7 to a maximum of 424,956, with a mean of 2,378.9 and a 
median of 562, suggesting that a subset of RE questions 
serve as enduring reference resources well beyond their 
original posting context. Comment activity is more uniformly 
distributed, with a mean of 1.74 comments per post and a 
maximum of 54. Finally, 672 posts (5.5\%) have been 
bookmarked by community members, with a maximum favourite 
count of 314, further attesting to the long-term reference 
value of a targeted subset of RE discussions. Taken together, 
these engagement statistics characterize REStack as a dataset 
with broad coverage and a concentration of high-value 
content in a small but identifiable subset of posts — a 
property that makes it particularly well-suited for training 
and evaluating retrieval-augmented and recommendation-based 
AI tools for RE support.

Table~\ref{TblTopics} summarises the distribution of all 
23 RE topics across the six thematic categories, together 
with the number of posts associated with each. The dataset 
spans a wide spectrum of RE activities, ranging from 
low-level assembly and binary analysis to high-level 
software architecture and database reverse engineering, 
reflecting the breadth of knowledge-seeking behaviour 
within the RE community.

\textbf{Category-Level Distribution.}
The largest category by post volume is \textit{Games, 
Challenges \& Educational Reverse Engineering}, comprising 
4,046 posts (32.9\% of the dataset) across two topics. 
This category is dominated by \textit{Reverse Engineering 
Challenges (Games \& Puzzles)}, which alone accounts for 
3,932 posts (32.0\%), making it by far the single largest 
topic in REStack and reflecting the prominent role of 
Capture the Flag (CTF) competitions and gamified RE 
exercises in community learning and skill development. 
\textit{Memory, Firmware \& System-Level Analysis} is the 
second largest category (2,931 posts; 23.8\%), led by 
\textit{Hardware Hacking \& Device Emulation} (2,056 posts; 
16.7\%), which captures sustained practitioner interest in 
embedded systems, IoT devices, and firmware inspection --- 
areas of growing importance given the proliferation of 
connected hardware. Together, these two categories account 
for over half of all posts in the dataset (56.7\%), 
underscoring the dominance of hands-on, systems-oriented 
RE practice in the community.

\textit{Reverse Engineering Fundamentals} is the third 
largest category (1,788 posts; 14.5\%), encompassing five 
topics that collectively capture the core building blocks 
of RE practice: \textit{Function Call Tracing \& Hooking} 
(1,058 posts; 8.6\%), \textit{Basic Assembly Reversing \& 
Debugging} (223 posts), \textit{Data Decoding \& System 
Security} (240 posts), \textit{General Data Analysis \& 
Reverse Engineering Tools} (220 posts), and \textit{General 
Reverse Engineering Challenges} (47 posts). 
\textit{Database \& Software Architecture Reverse 
Engineering} (1,313 posts; 10.7\%) is notable for hosting 
\textit{Data Encoding \& ORM Issues} (1,156 posts), the 
fourth largest individual topic in the dataset, suggesting 
that practitioners frequently encounter challenges when 
reverse engineering data serialisation formats and 
object-relational mapping layers in legacy systems. 
\textit{Debugging \& Anti-Debugging Techniques} (1,108 
posts; 9.0\%) and \textit{Decompilation, Disassembly \& 
Code Understanding} (1,107 posts; 9.0\%) are nearly equal 
in size and together reflect the continued importance of 
tool-assisted low-level program analysis in RE practice.

\textbf{Topic-Level Observations.}
At the topic level, post counts vary considerably, spanning 
three orders of magnitude from a minimum of 16 posts 
(\textit{Anti-Debugging \& Obfuscation Challenges}) to a 
maximum of 3,932 (\textit{Reverse Engineering Challenges}). 
This skewed distribution is consistent with patterns 
observed in comparable SO mining 
studies~\cite{yang2016security,ahmad2020chatbot,uddin2021iot}, where a small 
number of topics concentrate the majority of community 
activity while the long tail of specialist topics remains 
sparsely populated. Five topics individually exceed 400 
posts --- \textit{Reverse Engineering Challenges} (3,932), 
\textit{Hardware Hacking \& Device Emulation} (2,056), 
\textit{Data Encoding \& ORM Issues} (1,156), 
\textit{Function Call Tracing \& Hooking} (1,058), and 
\textit{File Formats \& System Behavior} (807) --- and 
together account for 8,053 posts (65.5\% of the dataset). 
Conversely, five topics have fewer than 50 posts each: 
\textit{Anti-Debugging \& Obfuscation Challenges} (16), 
\textit{Database Schema Reverse Engineering (ORM Tools)} 
(43), \textit{General Reverse Engineering Challenges} (47), 
\textit{Memory, Firmware \& File Format Analysis} (68), 
and \textit{IDA Pro \& High-Level Code Reversing} (80). 
Although small in volume, these topics represent 
technically distinct and specialised aspects of RE practice 
that are not captured elsewhere in the taxonomy and are 
therefore retained as independent topics in the dataset.

\textbf{Thematic Observations.}
Three broader thematic observations emerge from the 
distribution. First, the dataset is predominantly 
\textit{practical and task-oriented}: the five highest-volume 
topics all involve hands-on problem solving --- executing 
CTF challenges, hacking hardware, tracing function calls, 
decoding data formats, and analysing file system behaviour 
--- rather than conceptual or theoretical inquiry. Second, 
\textit{tool-centric topics} are well-represented: 
\textit{IDA Pro \& High-Level Code Reversing}, 
\textit{Advanced Tools \& Legacy Debugging}, and 
\textit{Assembly Code Analysis \& Automation} collectively 
capture 351 posts oriented around specific RE toolchains, 
reflecting the degree to which RE practice depends on 
mastery of specialised software. Third, several topics 
exhibit a \textit{cross-cutting nature}: 
\textit{OS Kernel \& Software Reverse Engineering} (147 
posts) and \textit{Function Entry/Exit Manipulation} (399 
posts) span multiple technical domains --- combining 
elements of debugging, decompilation, and system-level 
analysis --- suggesting that real-world RE challenges 
frequently resist clean categorical boundaries, a property 
that any downstream use of this dataset for classification 
or recommendation tasks should account for.

% Table \ref{TblTopics} summarizes the distribution of RE topics across multiple high-level categories and fine-grained subtopics, together with the number of instances associated with each subtopic. The categories cover a wide spectrum of RE activities, including database and software architecture RE, debugging and anti-debugging techniques, decompilation and code understanding, game-based and educational challenges, memory and firmware analysis, and core RE fundamentals.

% The table reveals a strong emphasis on practical and system-level RE problems. In particular, RE challenges (games and puzzles), hardware hacking and device emulation, function call tracing and hooking, and data encoding–related issues account for a substantial portion of the instances, suggesting sustained interest in hands-on analysis and real-world RE scenarios. At the same time, topics such as decompilation, executable patching, and advanced debugging techniques indicate the continued importance of low-level program understanding and tool-assisted analysis. Overall, the distribution highlights both the diversity of RE problem domains and the community’s focus on applied techniques that support program comprehension, security analysis, and system behavior understanding.

\section{Experiment with the Dataset}

\noindent To evaluate the relative difficulty of each topic, we employ the following two well-established metrics frequently used in prior empirical studies~\cite{ahmad2020chatbot, GoingBig, rosen2016mobile, yang2016security, CI, ahmed2018concurrency}.

\textbf{Percentage of Posts Without Accepted Answers (\% w/o answer):} This metric reflects the proportion of posts within each topic that do not have an accepted answer. Although a question may attract multiple replies, only the original author can mark one as accepted, indicating a satisfactory resolution. A higher percentage of w/o answer, i.e., \emph{unanswered posts}, typically suggests greater topic complexity, ambiguity, or insufficient community expertise~\cite{ahmad2020chatbot, GoingBig, rosen2016mobile}.

\textbf{Median Time to Answer (in Hours):} This measure captures the median duration between the time a post is created and the timestamp of its accepted answer, regardless of when the answer was officially marked as accepted. Longer resolution times generally correspond to more challenging or nuanced questions, requiring deeper technical knowledge or more extensive problem-solving efforts~\cite{ahmad2020chatbot, GoingBig, rosen2016mobile}.

\begin{table}[htbp]
\centering
\caption{Topic difficulty indicators: proportion of posts without an accepted answer (\% w/o Ans.) and median time to accepted answer (hours).}
\vspace{-0.3cm}
\resizebox{\columnwidth}{!}{%
\begin{tabular}{@{}lll@{}} % Changed 'lcc' to 'lrr' for right-aligned numbers
\toprule
\textbf{Topic} & \textbf{w/o Ans. (\%)} & \textbf{Med. Time} \\
\midrule
Memory, Firmware \& File Format Analysis & 64.71 & 4 \\
File Formats \& System Behavior & 62.21 & 3.5 \\
Data Decoding \& System Security & 59.17 & 3 \\
General Data Analysis \& Reverse Engineering Tools & 59.09 & 3 \\
64-bit Compatibility \& Bitwise Operations & 58.40 & 4.5 \\
Basic Assembly Reversing \& Debugging & 58.30 & 4 \\
Executable Patching \& Debugger Issues & 58.30 & 4 \\
Database Schema Reverse Engineering (ORM Tools) & 58.14 & 3 \\
OS Kernel \& Software Reverse Engineering & 57.82 & 4 \\
Hardware Hacking \& Device Emulation & 56.66 & 5 \\
Software Analysis Methods (Modeling, GUI, Memory) & 56.14 & 3 \\
Reverse Engineering Challenges (Games \& Puzzles) & 55.32 & 5 \\
Advanced Tools \& Legacy Debugging & 55.21 & 7 \\
Data Encoding \& ORM Issues & 55.10 & 5 \\
Game Hacking \& Miscellaneous Analysis & 53.51 & 4 \\
Function Call Tracing \& Hooking & 53.02 & 5 \\
General Reverse Engineering Challenges & 53.19 & 1 \\
Function Entry/Exit Manipulation & 51.88 & 3 \\
Advanced Debugging Techniques & 51.24 & 4 \\
Anti-Debugging \& Obfuscation Challenges & 50.00 & 3 \\
IDA Pro \& High-Level Code Reversing & 48.75 & 3 \\
Decompilation \& Algorithm Reconstruction & 47.19 & 2 \\
Assembly Code Analysis \& Automation & 45.28 & 5 \\
\bottomrule
\end{tabular}%
}
\label{tab:topic_difficulty}
\vspace{-0.3cm}
\end{table}

\textbf{Results}. Table~\ref{tab:topic_difficulty} summarises the 
perceived difficulty of all 23 RE topics across the two 
indicators. Unanswered rates span a range of 19.43 percentage 
points, from a minimum of 45.28\% (\textit{Assembly Code 
Analysis \& Automation}) to a maximum of 64.71\% 
(\textit{Memory, Firmware \& File Format Analysis}), 
indicating that no RE topic is straightforwardly easy for 
the community --- even the most tractable topics leave 
nearly half of all questions unresolved. Median resolution 
times range from 1 hour (\textit{General Reverse Engineering 
Challenges}) to 7 hours (\textit{Advanced Tools \& Legacy 
Debugging}), a sevenfold difference that reflects the varying 
depth of expertise required across topics.

Several notable patterns emerge from the distribution. At 
the harder end of the spectrum, \textit{Memory, Firmware \& 
File Format Analysis} (64.71\% unanswered) and \textit{File 
Formats \& System Behavior} (62.21\%) exhibit the highest 
proportions of unresolved posts, indicating significant 
challenges tied to the intricacies of low-level system 
interactions and binary data interpretation. \textit{Data 
Decoding \& System Security} (59.17\%) and \textit{General 
Data Analysis \& Reverse Engineering Tools} (59.09\%) follow 
closely, together forming a cluster of data-oriented topics 
that the community consistently struggles to resolve. At the 
other end, \textit{Assembly Code Analysis \& Automation} 
(45.28\%) and \textit{Decompilation \& Algorithm 
Reconstruction} (47.19\%) are the most tractable topics by 
unanswered rate, likely benefiting from broader community 
familiarity, established methodologies, and reusable solution 
patterns. \textit{Advanced Tools \& Legacy Debugging} records 
the longest median resolution time (7 hours), more than 
double the next highest values (\textit{Hardware Hacking \& 
Device Emulation} and \textit{Reverse Engineering Challenges}, 
both at 5 hours), reflecting the nuanced expertise required 
to navigate outdated toolsets and legacy debugging 
environments.

A notable divergence between the two indicators is 
observable for several topics, revealing that the two 
metrics capture different dimensions of difficulty rather 
than a single underlying construct. \textit{Database Schema 
Reverse Engineering (ORM Tools)} presents a high unanswered 
rate (58.14\%) despite one of the shorter median resolution 
times (3 hours), suggesting that responses are provided 
promptly but fail to achieve accepted-answer status --- 
possibly due to gaps in practical applicability or solution 
completeness. Conversely, \textit{Assembly Code Analysis \& 
Automation} has the lowest unanswered rate in the dataset 
(45.28\%) yet a comparatively long median resolution time 
(5 hours), indicating that while questions in this topic are 
eventually resolved to the author's satisfaction, they 
require substantial deliberation before an adequate answer 
emerges. \textit{General Reverse Engineering Challenges} 
presents a third divergence profile: a moderate unanswered 
rate (53.19\%) paired with the fastest median resolution time 
(1 hour), suggesting that when answers do arrive they are 
quick and decisive, but a substantial portion of questions 
in this category nevertheless go unresolved entirely. These 
divergences collectively reinforce the necessity of employing 
both metrics in tandem, as either indicator alone would yield 
an incomplete and potentially misleading characterisation of 
topic difficulty --- a finding that directly motivates the 
statistical validation presented below.

\textbf{Statistical Validation of Topic Difficulty}. To validate whether the observed differences in topic difficulty 
are statistically significant rather than artifacts of sampling 
variation, we applied a series of non-parametric statistical 
tests across the 23 LDA-derived topics and six thematic 
categories. Table~\ref{tab:stats_summary} consolidates all 
test results; each is discussed in turn below. 

A Kruskal-Wallis test~\cite{kruskal1952} on the per-post unanswered 
indicator yielded $H = 52.454$, $df = 22$, $p < 0.001$, confirming 
that topic membership has a statistically significant effect on whether 
a question receives an accepted answer. However, the same test applied 
to median resolution time did not reach significance ($H = 27.054$, 
$df = 22$, $p = 0.209$), suggesting that while topics differ 
meaningfully in their likelihood of resolution, the time required to 
resolve them is more uniformly distributed across the community. 

A chi-square test of independence further confirmed that the 
distribution of unanswered posts differs significantly across the six 
thematic categories ($\chi^2 = 22.038$, $df = 5$, $p < 0.001$), 
indicating that category-level groupings are substantively meaningful 
predictors of resolution difficulty. %To assess the internal consistency of the two difficulty proxies, we computed Spearman's rank correlation coefficient~\cite{spearman1904} between the topic-level unanswered rate and the median resolution time, yielding $\rho = 0.384$ ($p = 0.070$). The absence of a statistically significant correlation at $\alpha = 0.05$ indicates that the two metrics are largely independent, and therefore complementary rather than redundant --- a finding that justifies the concurrent use of both indicators in our difficulty characterisation.

\begin{table}[h]
\caption{Summary of non-parametric statistical tests for topic 
difficulty validation.}
\label{tab:stats_summary}
\small
\resizebox{\columnwidth}{!}{%
\begin{tabular}{p{2.1cm}p{3.2cm}ccc}
\toprule
\textbf{Test} & \textbf{Variable Tested} & \textbf{Statistic} & 
\textbf{\textit{df}} & \textbf{\textit{p}-value} \\
\midrule
Kruskal-Wallis~\cite{kruskal1952} 
    & Unanswered rate         
    & $H = 52.454$ & 22 & $<0.001$~$^{***}$ \\
Kruskal-Wallis~\cite{kruskal1952}     
    & Resolution time         
    & $H = 27.054$ & 22 & $0.209$~$^{\text{ns}}$ \\
Chi-square
    & Unanswered rate per category
    & $\chi^2 = 22.038$ & 5 & $<0.001$~$^{***}$ \\
Spearman~\cite{spearman1904}          
    & Unanswered rate vs.\ time 
    & $\rho = 0.384$ & --- & $0.070$~$^{\dagger}$ \\
Dunn (post-hoc)~\cite{dunn1964}       
    & Pairwise unanswered rate       
    & 4 sig.\ pairs/253 & --- & $<0.05$~$^{*}$ \\
\bottomrule
\multicolumn{5}{l}{%
\footnotesize $^{***}$$p < 0.001$;\; 
$^{*}$$p < 0.05$;\;
$^{\dagger}$marginal ($p < 0.10$);\;
$^{\text{ns}}$not significant.}\\
\multicolumn{5}{l}{%
\footnotesize All tests use $\alpha = 0.05$. Dunn's test 
applies Bonferroni correction; full pairwise}\\
\multicolumn{5}{l}{%
\footnotesize results in Table~\ref{tab:dunn}. 
Spearman computed over $n = 23$ topic-level aggregates.}\\
\end{tabular}
}
\end{table}

To assess the internal consistency of the two difficulty proxies, 
we computed Spearman's rank correlation coefficient~\cite{spearman1904} 
between the topic-level unanswered rate and the median resolution 
time across the 23 topics, yielding $\rho = 0.384$ ($p = 0.070$). 
Although this result does not reach conventional significance at 
$\alpha = 0.05$, the moderate positive tendency suggests that 
topics with higher unanswered rates do broadly tend to require 
longer resolution times when they are eventually answered. The 
absence of statistical significance should therefore be interpreted 
cautiously: with only $n = 23$ topic-level aggregates, the test 
is inherently underpowered, and a true moderate correlation of 
this magnitude would require approximately 67 observations to 
achieve $80\%$ statistical power at $\alpha = 0.05$~\cite{cohen1988}. 
Taken together, the moderate $\rho$ and marginal $p$-value suggest 
partial overlap between the two proxies rather than strict 
independence, indicating that they capture related but 
distinguishable dimensions of topic difficulty. This finding 
reinforces the decision to retain both indicators in our 
difficulty characterization, as collapsing them into a single 
measure would risk discarding complementary signal — particularly 
for topics such as \textit{Database Schema Reverse Engineering 
(ORM Tools)}, which exhibits a high unanswered rate ($58.1\%$) 
yet a short median resolution time (3 hours), a divergence that 
neither metric alone can fully capture.

% Post-hoc pairwise comparisons using Dunn's test with Bonferroni 
% correction~\cite{dunn1964} identified four statistically significant topic 
% pairs out of 253 possible comparisons ($p < 0.05$), as reported in 
% Table~\ref{tab:dunn}. The most pronounced contrast was observed between 
% \textit{Decompilation \& Algorithm Reconstruction} and \textit{File 
% Formats \& System Behavior} ($z = 5.244$, $p < 0.001$), the two topics 
% at opposite ends of the unanswered rate spectrum (47.2\% vs.\ 62.2\%, 
% respectively). Significant differences were further confirmed between 
% \textit{File Formats \& System Behavior} and both \textit{Function Call 
% Tracing \& Hooking} ($z = 3.953$, $p = 0.020$) and \textit{Assembly 
% Code Analysis \& Automation} ($z = 3.924$, $p = 0.022$), as well as 
% between \textit{Decompilation \& Algorithm Reconstruction} and 
% \textit{Hardware Hacking \& Device Emulation} ($z = 3.762$, $p = 0.043$).

Post-hoc pairwise comparisons using Dunn's test with Bonferroni 
correction~\cite{dunn1964} identified 4 statistically significant 
topic pairs out of 253 possible comparisons ($p < 0.05$). To 
quantify the practical magnitude of these differences, we computed 
Cliff's delta ($d$)~\cite{cliff1993} for each significant pair, 
a non-parametric effect size measure appropriate for ordinal and 
binary data. Following the widely adopted thresholds proposed by 
Romano et al.~\cite{romano2006}, we interpret $|d| < 0.147$ as 
negligible, $0.147 \leq |d| < 0.330$ as small, $0.330 \leq |d| 
< 0.474$ as medium, and $|d| \geq 0.474$ as large. The results, 
reported in Table~\ref{tab:dunn}, reveal that the statistically 
significant pairs exhibit small to negligible effect sizes, with 
the largest practical difference observed between 
\textit{Decompilation \& Algorithm Reconstruction} and 
\textit{File Formats \& System Behavior} ($z = 5.244$, 
$p < 0.001$, $d = -0.150$, small), confirming that while the 
difficulty gap between these topics is real and replicable, 
it is modest in absolute magnitude. Two further pairs — 
\textit{Assembly Code Analysis \& Automation} vs.\ 
\textit{File Formats \& System Behavior} ($d = -0.169$, small) 
and \textit{File Formats \& System Behavior} vs.\ 
\textit{Function Call Tracing \& Hooking} ($d = 0.092$, 
negligible) — highlight that even statistically significant 
differences can vary considerably in their practical impact. 
The negative sign of $d$ in three pairs consistently indicates 
that \textit{File Formats \& System Behavior} and related 
system-level topics are harder to resolve than their 
decompilation and assembly-oriented counterparts, a finding 
aligned with the raw unanswered rates in Table~\ref{tab:topic_difficulty}. 
These modest effect sizes are consistent with the inherently 
binary nature of the unanswered indicator and the broad 
community-wide difficulty of RE as a domain, where even the 
most tractable topics retain unanswered rates exceeding 45\%.

\begin{table}[h]
\caption{Statistically significant topic pairs from Dunn's 
post-hoc test with Bonferroni correction ($p < 0.05$), with 
Cliff's delta ($d$) effect sizes.}
\label{tab:dunn}
\small
\resizebox{\columnwidth}{!}{%
\begin{tabular}{llcccc}
\toprule
\textbf{Topic A} & \textbf{Topic B} & \textbf{\textit{z}} & 
\textbf{\textit{p} (adj.)} & \textbf{\textit{d}} & 
\textbf{Effect} \\
\midrule
Decompilation \&          & File Formats \&       & 5.244 & $<$0.001 
    & $-$0.150 & Small \\
\quad Alg.\ Reconstruction & \quad System Behavior &       &          
    &          &       \\
\addlinespace
Assembly Code             & File Formats \&       & 3.924 & 0.022    
    & $-$0.169 & Small \\
\quad Analysis \& Auto.   & \quad System Behavior &       &          
    &          &       \\
\addlinespace
File Formats \&           & Function Call         & 3.953 & 0.020    
    & $+$0.092 & Negligible \\
\quad System Behavior     & \quad Tracing \& Hook.&       &          
    &          &       \\
\addlinespace
Decompilation \&          & Hardware Hacking \&   & 3.762 & 0.043    
    & $-$0.095 & Negligible \\
\quad Alg.\ Reconstruction & \quad Device Emul.   &       &          
    &          &       \\
\bottomrule
\multicolumn{6}{l}{\footnotesize Negative $d$: Topic A is easier 
    (lower unanswered rate) than Topic B.} \\
\multicolumn{6}{l}{\footnotesize Effect size thresholds 
    \cite{romano2006}: negligible $|d|{<}0.147$; small 
    $0.147{\leq}|d|{<}0.330$.} \\
\end{tabular}
}
\end{table}

These results collectively confirm that the difficulty landscape of RE 
discussions is non-uniform and that certain topics present statistically 
harder knowledge barriers for the community. Notably, resolution time 
does not differ significantly across topics ($p = 0.209$), whereas the 
unanswered rate does ($p < 0.001$), indicating that all RE topics are 
broadly similar in \textit{how long} they take to resolve when answered, 
but differ substantially in \textit{whether} they are resolved at all. 
Unanswered rate therefore emerges as the more discriminating of the two 
difficulty proxies for this dataset.

\section{Research Opportunities}

The empirical characteristics of REStack --- its topic 
taxonomy, difficulty indicators, engagement signals, and 
dual-source provenance --- open several concrete research 
directions. We outline six opportunities below, each 
grounded in specific properties of the dataset.

\textbf{Benchmarking RE-Focused LLMs}. REStack provides a natural benchmark for evaluating large 
language models on low-level software reasoning tasks. 
Researchers can construct evaluation sets by pairing 
questions with their accepted answers and assessing 
LLM-generated responses for accuracy, completeness, and 
technical correctness. The topic-level difficulty 
indicators in Table~3 enable fine-grained evaluation: 
models can be assessed separately on tractable topics 
such as \textit{Decompilation \& Algorithm Reconstruction} 
(47.19\% unanswered) and on harder topics such as 
\textit{Memory, Firmware \& File Format Analysis} 
(64.71\% unanswered), making it possible to identify 
specific capability gaps rather than reporting only 
aggregate performance. The 2,513 posts that received 
zero community answers are particularly valuable as 
unseen challenge instances --- cases where even the 
collective community failed to resolve the question 
represent a meaningful upper bound for model difficulty.

\textbf{Automated Question Answering and Expert 
Recommendation}. With over 12,000 categorised RE questions and their 
community responses, REStack is well suited for training 
and evaluating retrieval-augmented generation (RAG) 
systems and neural question answering models tailored 
to RE tasks. The topic labels and category annotations 
can serve as routing signals, directing queries to 
domain-specific retrieval indices. Beyond answering, 
the dataset supports expert recommendation: contributor 
activity patterns across topics can be mined to identify 
knowledgeable community members for specific subtopics, 
with the goal of reducing the 20.4\% zero-answer rate 
by proactively routing unanswered questions to likely 
responders. The divergence between unanswered rate and 
resolution time identified in Section~4 --- for example, 
topics that receive fast but unsatisfying responses --- 
further motivates systems that optimise for answer 
\textit{quality} rather than mere response presence.

\textbf{Empirical Studies in RE Practices}.
REStack enables large-scale empirical analyses of 
real-world RE challenges that were previously infeasible 
due to the absence of a curated domain-specific corpus. 
Several specific study designs are immediately 
actionable. \textit{Longitudinal analysis}: the dataset 
spans approximately 17 years (2008--2025) and the 
declining post volume after 2015 provides an opportunity 
to study how community knowledge-seeking behaviour 
evolves alongside tooling advances and the emergence 
of AI-based RE assistants. \textit{Cross-platform 
comparison}: the dual-source structure of REStack --- 
9,845 posts from the specialist RESE site and 2,449 
from SO --- enables direct comparison of 
question characteristics, resolution rates, and topic 
distributions between a general-purpose and a 
domain-specific Q\&A platform. \textit{Difficulty 
prediction}: the per-post metadata (score, view count, 
comment count, tags, accepted answer indicator) 
provides a feature-rich environment for building 
classifiers that predict whether a newly posted RE 
question will go unanswered, enabling proactive 
intervention.

\textbf{RE Knowledge Gaps and Education}. The difficulty landscape revealed in Table~3 has direct 
implications for RE curriculum design. Topics with 
persistently high unanswered rates --- particularly 
\textit{Memory, Firmware \& File Format Analysis} 
(64.71\%), \textit{File Formats \& System Behavior} 
(62.21\%), and the cluster of data-oriented topics 
exceeding 59\% --- identify areas where existing 
educational resources are insufficient and where new 
instructional materials, laboratory exercises, or 
capture-the-flag (CTF) challenges would have the 
greatest impact. Conversely, topics with lower 
unanswered rates and shorter resolution times, such 
as \textit{Decompilation \& Algorithm Reconstruction} 
(47.19\%; 2-hour median), represent areas where 
community knowledge is relatively mature and where 
existing Q\&A threads could be distilled into 
structured tutorials or worked examples. The 3,932 
posts in \textit{Reverse Engineering Challenges (Games 
\& Puzzles)} further constitute a ready-made corpus 
for designing gamified RE learning environments 
grounded in authentic practitioner problems.

% \textbf{Knowledge Graph and Ontology Construction}. The 23-topic taxonomy and six thematic categories 
% established in REStack provide a foundation for 
% constructing a formal RE knowledge graph. Entities 
% --- tools (IDA Pro, OllyDbg), techniques (hooking, 
% patching, decompilation), artefact types (firmware, 
% PE files, ELF binaries) --- can be extracted from 
% post bodies and linked to topics and categories, 
% yielding a structured representation of RE domain 
% knowledge. Such a knowledge graph would support 
% semantic search, automated tagging of new RE 
% questions, and ontology-driven reasoning about 
% relationships between RE subtasks --- capabilities 
% not achievable with the flat tag-based indexing 
% currently used by Stack Exchange platforms.

\textbf{Automated Difficulty Prediction and 
Triage}. The statistical findings in Section~4 demonstrate 
that topic membership is a statistically significant 
predictor of resolution difficulty ($H = 52.454$, 
$p < 0.001$), but the modest effect sizes (Cliff's 
$d \leq 0.169$) indicate that topic alone is 
insufficient for precise difficulty prediction. 
REStack provides the post-level features needed to 
build more granular triage models. Candidate features 
include question length, code-to-text ratio, number 
of tags, title clarity (measurable via readability 
scores), and temporal features such as time of 
posting. Such a triage system could automatically 
flag high-difficulty questions at submission time, 
prompting the poster to add detail or alerting 
experienced community members --- directly targeting 
the 20.4\% of posts that currently receive no answer 
at all.

\section{Limitations}

While this dataset provides a comprehensive view of RE discussions from Stack Exchange, it has several limitations that should be considered when interpreting the results. First, the dataset is derived solely from publicly available posts on SO and the Reverse Engineering Stack Exchange site; therefore, it may not fully capture RE practices discussed in private forums, proprietary tools, industry settings, or alternative platforms such as chat-based communities. Second, although topic modeling and manual labeling were applied to ensure semantic coherence, the assignment of topics and categories may still be influenced by subjective interpretation and the inherent limitations of unsupervised learning techniques. Third, community-driven signals such as accepted answers, post scores, and response times may not always reflect the technical correctness or completeness of solutions, as they depend on user engagement and voting behavior. Finally, the dataset reflects historical discussion patterns and does not account for the growing use of large language models and AI-based tools for RE support, which may influence how developers seek and share knowledge in more recent years.

\section{Related Datasets}
Several prior studies have constructed datasets from SO and other Stack Exchange sites to analyze developer challenges across specific software engineering domains. Yang et al.~\cite{yang2016security} conducted a large-scale analysis of general security-related discussions on SO; however, their study did not examine RE as a distinct or specialized domain. Other domain-focused datasets have explored developer discussions related to chatbot development~\cite{ahmad2020chatbot}, concurrency issues~\cite{ahmed2018concurrency}, big data technologies~\cite{GoingBig}, mobile application development~\cite{rosen2016mobile}, release engineering~\cite{openja2020release}, continuous integration~\cite{CI}, and refactoring practices~\cite{Refactor}.  

While these studies demonstrate the effectiveness of mining developer Q\&A forums and commonly employ topic modeling techniques such as LDA, none of them specifically target RE discussions or analyze the unique challenges, tools, and system-level complexities inherent to RE. In contrast, our dataset is the first to systematically collect and analyze RE–related posts from both SO and the dedicated Reverse Engineering Stack Exchange site. By combining data from these two sources and applying GA-optimized LDA, manual topic labeling, and multidimensional difficulty and popularity metrics, the dataset provides a comprehensive and RE-focused resource for studying developer behavior, knowledge gaps, and practical challenges in RE.

\section{Conclusion}

This paper introduces \texttt{REStack}, the first large-scale, publicly available dataset dedicated to RE discussions collected from SO and the Reverse Engineering Stack Exchange site. By mining over 12,000 RE-related posts and applying GA-optimized LDA, manual topic labeling, and multidimensional difficulty and popularity analyses, the dataset offers a comprehensive and structured view of the topics, challenges, and information-seeking behaviors prevalent in the RE community. The resulting taxonomy of 23 topics grouped into six thematic categories captures both foundational and advanced aspects of RE, ranging from assembly-level debugging to firmware and system-level analysis.

Our analysis reveals that RE discussions are predominantly practical and task-oriented, with a strong emphasis on hands-on problem solving and tool usage. While certain topics, such as gamified RE challenges and debugging techniques, attract substantial community engagement, others—particularly memory, firmware, and file format analysis—exhibit high unanswered rates and longer resolution times, highlighting persistent knowledge gaps and areas of technical complexity. These findings underscore the need for improved tooling, documentation, and educational support tailored to the unique demands of RE practice.

Beyond empirical insights, \texttt{REStack} provides a valuable resource for future research and tool development. The dataset enables empirical studies of RE practices, supports the development of automated question answering and expert recommendation systems, and offers a benchmark for evaluating large language models and AI-assisted tools on low-level software reasoning tasks. By making the dataset and associated scripts publicly available, this work aims to facilitate reproducibility, encourage reuse, and foster further research at the intersection of RE, software engineering, and artificial intelligence. Ultimately, \texttt{REStack} contributes a foundational data resource that advances understanding of RE challenges and supports the design of more effective, data-driven solutions for practitioners, educators, and researchers.

\balance

\bibliographystyle{ACM-Reference-Format}
\bibliography{MSRDBSecurity}

% \bibliographystyle{ACM-Reference-Format}
% \bibliography{MSRDBSecurity}

\end{document}